\documentclass[aps,prl,twocolumn,groupedaddress,longbibliography,nobibnotes]{revtex4-1}
\usepackage{amsmath}
\usepackage{amssymb}
\usepackage{mathrsfs}
\usepackage[pdftex,colorlinks=true,urlcolor=blue,linkcolor=blue,citecolor=blue,breaklinks=true]{hyperref}
\usepackage{graphicx}
\usepackage{float}
\usepackage{dsfont}

\makeatletter
\def\l@subsection{%
 \l@@sections{section}{subsection}
}
\def\l@f@subsection{%
 \addpenalty{\@secpenalty}%
 \addvspace{0.5em plus\p@}%
}
\def\l@subsubsection#1#2{}
\makeatother

\begin{document}

\title{Critical Response of a Quantum van der Pol Oscillator}

\author{Shovan Dutta}
\email[E-mail: ]{sd843@cam.ac.uk}
\author{Nigel R. Cooper}
\email[E-mail: ]{nrc25@cam.ac.uk}
\affiliation{T.C.M. Group, Cavendish Laboratory, University of Cambridge, JJ Thomson Avenue, Cambridge CB3 0HE, United Kingdom\looseness=-1}

\date{\today}

\begin{abstract}
Classical dynamical systems close to a critical point are known to act as efficient sensors due to a strongly nonlinear response. We explore such systems in the quantum regime by modeling a driven van der Pol oscillator. We find the classical response survives down to one excitation quantum. At very weak drives, genuine quantum features arise, including diverging and negative susceptibilities. Further, the linear response is greatly enhanced by using a strong incoherent pump. These results are largely generic and can be probed in current experimental platforms suited for quantum sensing.
\end{abstract}

\maketitle

A key insight from the theory of phase transitions is that systems close to a critical point are highly sensitive to perturbations \cite{kardar2007statistical
}. This is evident, for example, in a diverging compressibility near a liquid-gas phase transition. Such divergences in susceptibility also occur in classical dynamical systems close to a bifurcation \cite{strogatz1994nonlinear}. A particularly important prototype is an oscillator with nonlinear damping, which can transition from a dormant to a limit-cycle phase across a Hopf bifurcation~\cite{marsden1976hopf}. Generically, the response of such a van der Pol (vdP) oscillator \cite{van1926lxxxviii
} to a resonant drive $\Omega$ grows as $\Omega^{1/3}$ at the critical point~\cite{duke2008critical}. This nonlinearity is what enables the ear and other biological sensors to detect very weak stimuli and maximally process environmental inputs \cite{duke2008critical, eguiluz2000essential, 
mora2011biological, munoz2018colloquium, roli2018dynamical}.

In this Letter, we ask if this increased sensitivity persists into the quantum regime. This could have important applications in the growing frontier of quantum sensing \cite{Degen2017, Braun2018}. Besides, understanding the dynamics of a quantum vdP oscillator is also of fundamental interest, as it represents an easy-to-implement nonequilibrium setting where coherent drive competes with incoherent dissipation. Such open quantum systems are at the forefront of modern physics \cite{Sieberer2016, Rotter2015}, led by advances in synthetic experimental platforms with unprecedented control and tunability \cite{Mueller2012, Houck2012, Aspelmeyer2014, Noh2016, Georgescu2014}.

At the semiclassical level, the vdP oscillator is realized in the physics of lasers \cite{Vahala2009, sargent1974laser}. Recent theoretical studies have also examined a quantum vdP oscillator, largely in the context of synchronization \cite{lee2013quantum, walter2014quantum, Loerch2016, Lee2014, Walter2014a, Ishibashi2017, Scarlatella2019}. However, its critical properties and response are as yet unexplored. Here, we fully characterize a driven quantum vdP oscillator by a standard master equation, finding surprising features which can be probed in experiments.

We find the classical response persists all the way down to one excitation quantum. At weaker drives, the quantum oscillator exhibits both divergent as well as negative susceptibilities. Further, in the limit-cycle phase, the linear response is only limited by two-body loss, providing a strong sensitivity enhancement over a passive system. Several of these features originate from a competition of energy scales and should generalize to other systems. After briefly reviewing the iconic classical vdP oscillator, we present results for the quantum response, concluding with a discussion of experimental realizations.

The classical vdP model describes a harmonic oscillator with nonlinear damping, $\ddot{x} = -\omega^2 x + (2\gamma_1- 8\gamma_2 x^2) \dot{x}$. Here, $\omega$ is the natural frequency, $\gamma_1$ is a negative damping which may arise from an energy source, and $\gamma_2$ is the leading nonlinear term which damps the system at large $x$. For $|\gamma_1| \ll \omega$, the small-amplitude oscillations are sinusoidal with a slowly-varying amplitude, $x = \text{Re}[\alpha(t) e^{{\rm i}\omega t}]$, where $\dot{\alpha} = \gamma_1\alpha - \gamma_2 |\alpha|^2 \alpha$ \cite{van1926lxxxviii}. Thus, if $\gamma_1<0$, the system is fully damped and all oscillations die out, whereas for $\gamma_1>0$, they reach a stable amplitude with $|\alpha|=\sqrt{\gamma_1/\gamma_2}$. When subjected to a resonant drive $F \cos(\omega t + \phi)$, the equation of motion gains a forcing term,
\begin{equation}
\dot{\alpha} = \gamma_1\alpha - \gamma_2 |\alpha|^2 \alpha + \Omega \;,
\label{classicaleom}
\end{equation}
where \smash{$\Omega = -{\rm i} e^{{\rm i}\phi} F/(2\omega)$}. We choose $\phi=\pi/2$ so that $\Omega$ is real and nonnegative.
From Eq.~\eqref{classicaleom}, the steady-state response at the critical point $\gamma_1=0$ is purely nonlinear, \smash{$\alpha = (\Omega/\gamma_2)^{1/3}$}. Hence, the susceptibility \smash{$\chi_{\text{cl}} \equiv \text{d}\alpha/\text{d}\Omega$} is divergent at zero drive, \smash{$\chi_{\text{cl}}\propto \Omega^{-2/3}$}. 
Away from criticality, the response is of the form $\alpha =\smash{\tilde{\Omega}^{1/3} f(\tilde{\gamma}_1/\tilde{\Omega}^{2/3})}$ where $\tilde{\gamma}_1 \equiv \gamma_1/\gamma_2$ and $\tilde{\Omega} \equiv \Omega/\gamma_2$ (details in Supplement~\footnote{See Supplemental Material for exact classical solution, analytic results for the quantum response at weak drives, mapping between density matrix and the Wigner function, and signal-to-noise ratio estimates
.}). In particular, the limiting behavior is such that
\begin{equation}
\alpha= 
   \begin{cases} 
      \sqrt{\gamma_1/\gamma_2} + \Omega/(2\gamma_1) & \text{for}\;\;\tilde{\gamma}_1 \gg \tilde{\Omega}^{2/3}\quad, \\
      \;\Omega/|\gamma_1| & \text{for}\;\;\tilde{\gamma}_1 \ll -\tilde{\Omega}^{2/3}\;,
   \end{cases}
\end{equation}
which are consistent with a diverging susceptibility, $\chi_{\text{cl}}\propto 1/|\gamma_1|$,  at the critical point $\gamma_1\to 0$.

We study a quantum version of the vdP model, where a quantum harmonic oscillator is subjected to several dissipative processes, as in Refs. \cite{lee2013quantum, walter2014quantum, Loerch2016, Lee2014, Walter2014a, Ishibashi2017, Scarlatella2019}. These are (i) one-particle loss with rate $\smash{\gamma_1^-}$, (ii) one-particle gain with rate $\smash{\gamma_1^+}$, and (iii) two-particle loss with rate~$\gamma_2$. As we will discuss later, such processes can be engineered in current experimental setups by coupling the oscillator to a suitable environment. Additionally, the oscillator is subject to a resonant drive of amplitude $\Omega$, given by the Hamiltonian $\hat{H} = \omega \hat{a}^{\dagger} \hat{a} + \Omega ({\rm i} e^{{\rm i}\omega t} \hat{a}^{\dagger} + \text{H.c.})$. Here, $\hat{a}$ annihilates a particle in the oscillator mode and $\hbar=1$.
It is convenient to work in the rotating frame where the Hamiltonian reads $\hat{H}={\rm i}\Omega(\hat{a}^{\dagger}-\hat{a})$. In general, the full dynamics of such a driven-dissipative system are governed by a master equation for the density matrix $\hat{\rho}$, obtained by tracing out the environment degrees of freedom \cite{Puri1977}. However, for typical atomic/photonic setups, the environment relaxes to equilibrium on optical timescales, several orders of magnitude faster than the system dynamics \cite{Daley2014}. Under such a routine Markov approximation \cite{Rivas2010, Rivas2014}, the master equation reduces to the Lindblad form
\begin{equation}
\dot{\hat{\rho}}= -{\rm i} [\hat{H},\hat{\rho}] + \gamma_1^+\hspace{0.03cm}\mathcal{D}[\hat{a}^{\dagger}]\hat{\rho}  + \gamma_1^-\hspace{0.03cm}\mathcal{D}[\hat{a}]\hat{\rho} + \gamma_2\hspace{0.03cm}\mathcal{D}[\hat{a}^2]\hat{\rho}\;,
\label{mastereq}
\end{equation}
where $\mathcal{D}[\hat{x}]\hat{\rho} \equiv \hat{x}\hat{\rho}\hat{x}^{\dagger} - \{\hat{x}^{\dagger}\hat{x},\hat{\rho}\}/2$ \cite{breuer2002theory, gardiner2004quantum}. Using Eq.~\eqref{mastereq}, or operator equations of motion \cite{sachdev1984atom}, the expectation value \smash{$\langle \hat{a} \rangle$} can be shown to obey
\begin{equation}
\langle\dot{\hat{a}}\rangle=[(\gamma_1^+ \hspace{-0.05cm}-\hspace{-0.05cm} \gamma_1^-)/2] \hspace{0.03cm}\langle\hat{a}\rangle - \gamma_2\hspace{0.03cm}\langle\hat{a}^{\dagger} \hat{a} \hat{a}\rangle + \Omega\;.
\label{modeq}
\end{equation}
Replacing $\hat{a}$ with a complex number $\alpha$ in Eq.~\eqref{modeq} leads to the classical limit in Eq.~\eqref{classicaleom} with $\gamma_1 = (\gamma_1^+ \hspace{-0.05cm}-\hspace{-0.03cm} \gamma_1^-)/2$. 
Hence, the critical case corresponds to $\gamma_1^+ = \gamma_1^-$. Note that although Eq.~\eqref{classicaleom} was derived in the small-amplitude limit, we will refer to Eq.~\eqref{mastereq} more generally as a quantum vdP oscillator, similar to Refs. \cite{lee2013quantum, walter2014quantum, Loerch2016, Lee2014, Walter2014a, Ishibashi2017, Scarlatella2019}.

We find the response by numerically solving Eq.~\eqref{mastereq} for the steady-state density matrix, then computing $\langle \hat{a} \rangle = \text{Tr}(\hat{a}\hat{\rho})$. It is most straightforward to use the Fock basis, $\hat{\rho} = \sum_{n n^{\prime}} \rho_{n,n^{\prime}} |n\rangle \langle n^{\prime}|$, where $n=0,1,2,\dots$ denotes the number of particles in the oscillator mode. The response is given by $\smash{\langle \hat{a} \rangle = \sum_n \sqrt{n} \rho_{n,n-1}}$. Past studies have found closed-form expressions for $\rho_{n,n^{\prime}}$ in special cases \cite{simaan1975quantum, dodonov1997exact, kheruntsyan1999wigner}.

\begin{figure}
\centering
\includegraphics[width=\columnwidth]{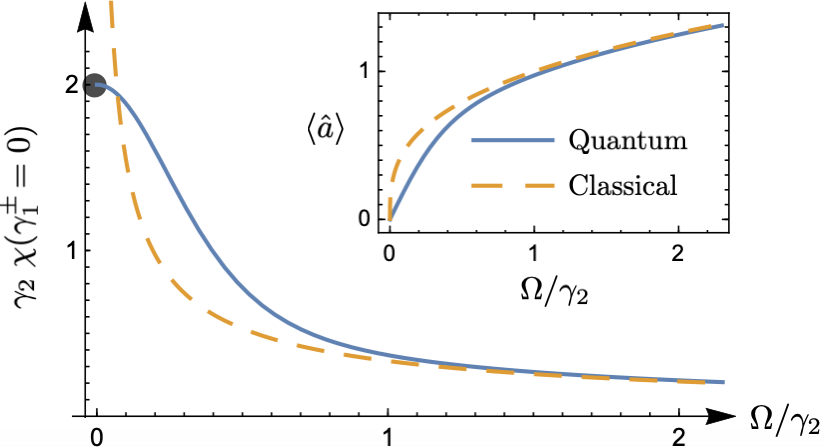}
\caption{\label{ss_critical}Susceptibility $\chi$ and response $\langle\hat{a}\rangle$ of a resonantly driven van der Pol oscillator at the critical point $\gamma_1^{\pm}=0$. Here, $\Omega$ is the drive amplitude and $\gamma_1^+$, $\gamma_1^-$, and $\gamma_2$ are the rates of one-particle gain, one-particle loss, and two-particle loss, respectively. The classical and quantum responses match down to $\langle\hat{a}\rangle\sim 1$, below which quantum fluctuations cut off the classical divergence, causing $\chi$ to saturate.
}
\end{figure}

First we consider the simplest critical case, $\gamma_1^+ = \gamma_1^- = 0$. Figure~\ref{ss_critical} shows the susceptibility $\chi \equiv \text{d} \langle \hat{a} \rangle/\text{d}\Omega$ as a function of drive. Note that it coincides with the classical result for $\langle \hat{a} \rangle \gtrsim 1$. However, at weaker drives, the classical divergence is cut off and $\chi$ saturates, producing a linear response.
We can understand this low-energy cutoff as follows. At $\Omega=0$, both $|0\rangle$ and $|1\rangle$ are steady states as there is only two-particle decay. A nonzero drive couples these neighboring Fock states, yielding $\rho_{10}, \rho_{21}\sim\Omega$ and $\rho_{22}\sim\Omega^2$. Hence, to linear order, one can restrict the dynamics to the lowest three levels, which gives $\langle \hat{a} \rangle = 2\Omega/\gamma_2$ (see Supplement \cite{Note1}). Stronger drives inject more particles, inducing a crossover to the classical limit. 

\begin{figure}
\centering
\includegraphics[width=\columnwidth]{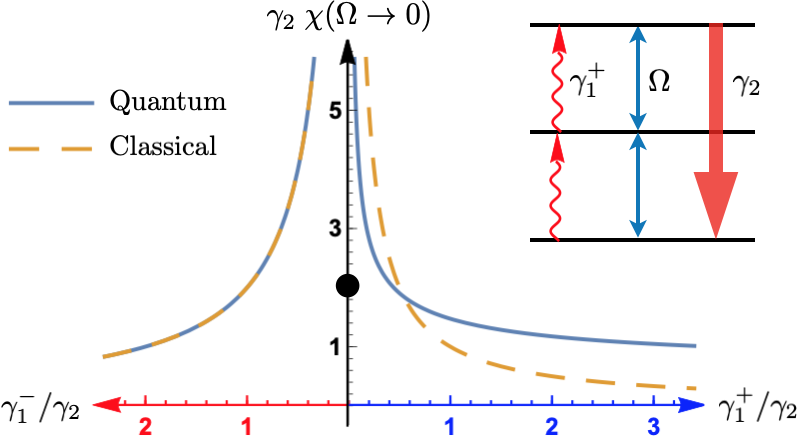}
\caption{\label{ss_zerofield}Zero-drive susceptibility as a function of damping $\gamma_1^-$ (with $\gamma_1^+=0$, red axis) and antidamping $\gamma_1^+$ (with $\gamma_1^-=0$, blue axis). Both classical and quantum susceptibilities diverge as $1/\gamma_1^{\pm}$, although at $\gamma_1^{\pm}=0$, the quantum susceptibility is finite (black circle, also see Fig.~\ref{ss_critical}). The inset shows transitions among the lowest three oscillator levels for $\gamma_1^+,\Omega \ll \gamma_2$ and $\gamma_1^-=0$. Since any two-particle excitation decays very rapidly, the dynamics are confined to this low-energy subspace.}
\end{figure}

We now consider departures from $\gamma_1^\pm= 0$ by allowing either 
$\gamma_1^+$ or $\gamma_1^-$ to be nonzero, spanning the transition from  quiescent ($\gamma_1^+=0$, $\gamma_1^-> 0$) to the limit-cycle regime ($\gamma_1^+>0$, $\gamma_1^-= 0$). In Fig.~\ref{ss_zerofield} we plot the zero-drive susceptibility across the transition, which, remarkably, shows a divergence as $1/\gamma_1^{\pm}$ at small damping or antidamping. This divergence can be understood from a competition of energy scales.
%
%
For $\gamma_1^+=0$ and $\gamma_1^->0$, the system is fully damped and the undriven steady state is $|0\rangle$. For weak drives $\Omega < \gamma_1^-$, the occupation remains small, so the two-particle decay is irrelevant. Thus, we can discard the nonlinear term in Eq.~\eqref{modeq}, which then reduces to a driven damped oscillator with steady state $\langle \hat{a} \rangle = 2\Omega/\gamma_1^-$, matching the classical result. The response for $\gamma_1^+>0$ and $\gamma_1^-=0$ is more involved, as the oscillator can have large occupations even at zero drive. In general, the linear response is given by Liouvillian perturbation theory \cite{li2014perturbative, albert2016geometry, konopik2018quantum}. However, for $\gamma_1^+\ll \gamma_2$, the dynamics are confined to the lowest three levels, as shown in Fig.~\ref{ss_zerofield} inset, which yields $\langle \hat{a} \rangle = 2\Omega/(9 \gamma_1^+)$ (see Supplement \cite{Note1}).

\begin{figure}
\centering
\includegraphics[width=\columnwidth]{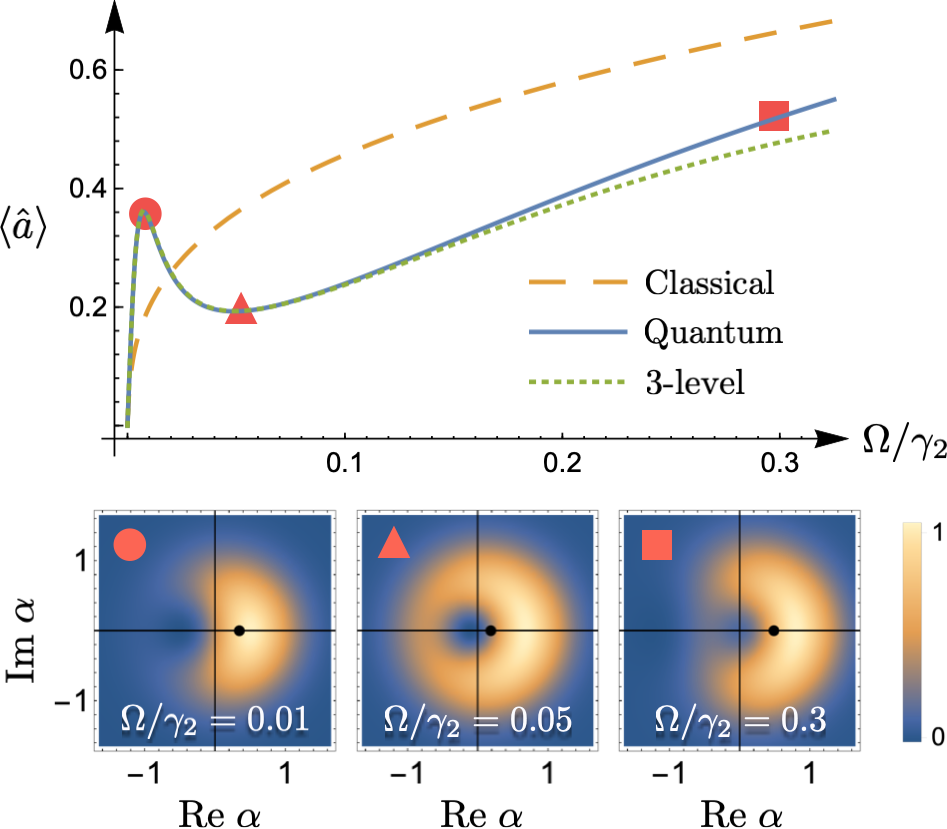}
\caption{\label{ss_negative}Top: response $\langle\hat{a}\rangle$ as a function of drive $\Omega$ for $\gamma_1^+=0$ and $\gamma_1^-/\gamma_2 = 0.02$. The quantum response is linear for $\Omega \lesssim \gamma_1^-$ and shows negative susceptibility for $\smash{\gamma_1^- \lesssim \Omega \lesssim (\gamma_1^-\gamma_2)^{1/2}}$. For $\Omega\gtrsim\gamma_2$, it agrees with the classical response, as in Fig.~\ref{ss_critical}. Bottom: rescaled Wigner functions $W(\alpha)$ describing a quasi-probability distribution in phase space \cite{Note1}. The response $\langle\hat{a}\rangle$ is given by the center of mass of $W(\alpha)$, shown by black dots.}
\end{figure}

At first sight, the results in Figs.~\ref{ss_critical} and \ref{ss_zerofield} might appear to be contradictory: at $\gamma_1^+=\gamma_1^-=0$, the zero-drive susceptibility is finite (black circle in Fig.~\ref{ss_critical}); however Fig.~\ref{ss_zerofield} shows that it diverges as the critical point is approached from either side. The resolution is that the susceptibility $\chi (\Omega\to 0, \gamma_1^{\pm} \to 0)$ depends on the order of the two limits. This is better understood by plotting the full response as a function of $\Omega$ for small $\gamma_1^{\pm}$, as shown in Fig.~\ref{ss_negative}. The response is strongly nonlinear for $\Omega \gtrsim \gamma_1^{\pm}$. Thus, although the linear susceptibility diverges, the linear region itself shrinks to zero as $\gamma_1^{\pm}\to~0$.

The full response exhibits four different regimes: (i) a linear response for $\smash{\Omega \lesssim \gamma_1^{\pm}}$, (ii) negative susceptibility for $\gamma_1^{\pm} \lesssim \Omega \lesssim (\gamma_1^{\pm} \gamma_2)^{1/2}$, (iii) an extended quantum response for $(\gamma_1^{\pm} \gamma_2)^{1/2} \lesssim \Omega \lesssim \gamma_2$, and (iv) classical response for $\smash{\Omega \gtrsim \gamma_2}$.
%
%
The first three regions are well reproduced by a three-level approximation, which yields, to lowest order in $\gamma_1^{\pm}/\gamma_2$ and $\Omega/\gamma_2$ (see Supplement \cite{Note1} for a derivation),
\begin{equation}
\langle\hat{a}\rangle \approx \frac{2\Omega}{\gamma_2} \left[\frac{(\gamma_1^+ +\gamma_1^-)\gamma_2 + 8 \Omega^2}{(3\gamma_1^+ + \gamma_1^-)^2 + 8 \Omega^2}\right].
\label{3levelres}
\end{equation}
Hence, in the linear region, $\chi \approx 2 (\gamma_1^+ \hspace{-0.04cm}+\hspace{-0.03cm}\gamma_1^-)/(3\gamma_1^+ \hspace{-0.04cm}+\hspace{-0.03cm} \gamma_1^-)^2$, which diverges for $\gamma_{\pm} \to 0$, whereas for $\Omega\hspace{-0.01cm}\sim\hspace{-0.01cm} (\gamma_1^{\pm}\gamma_2)^{1/2}$, $\langle\hat{a}\rangle \approx 2\Omega/\gamma_2 + (\gamma_1^+ \hspace{-0.05cm}+\hspace{-0.02cm} \gamma_1^-)/(4\Omega)$, which exhibits a local minimum. This nonmonotonic response is also evident in the Wigner functions shown in the lower panel of Fig.~\ref{ss_negative}, which describe a quasiprobability distribution in phase space \cite{gardiner2004quantum}. The Wigner function is widely used as an integral representation of the density matrix \cite{Note1}.

Physically, the separate regions originate from an interplay between drive and dissipation within a few-level manifold. Consider the case $\smash{\gamma_1^+=0}$ and $\smash{0<\gamma_1^- \ll \gamma_2}$. A weak drive couples the steady state $|0\rangle$ to $|1\rangle$, producing a coherence $\rho_{10} \sim \Omega/\gamma_1^-$. However, $\rho_{11}$ also grows with $\Omega$ and saturates at $1/2$. This saturation acts as a negative feedback for $\rho_{10}$, as seen from the equation of motion $\dot{\rho}_{11}=2\hspace{0.03cm}\Omega\hspace{0.03cm}\rho_{10}-\gamma_1^- \rho_{11}$. For $\rho_{11}\approx 1/2$, $\rho_{10}$ falls off as $\gamma_1^-/(4\Omega)$, giving rise to negative susceptibility. At larger drives, higher-energy modes become accessible and coherences can grow again. Then the response is similar to the case $\gamma_1^{\pm}=0$, with a purely quantum regime for $\langle\hat{a}\rangle \lesssim 1$ and a classical regime for $\langle\hat{a}\rangle \gtrsim 1$ (cf. Fig.~\ref{ss_critical}). A similar variation with drive is found for $\smash{0<\gamma_1^+ \ll \gamma_2}$.

Note that the concomitant divergent and negative susceptibilities are genuine quantum features which result from a very general set of conditions involving a competition between coherent and incoherent energy scales. Thus, we expect them to show up generically in driven-dissipative systems. For the vdP oscillator, we find they are robust to anharmonicity and detuning.

We have shown the quantum system exhibits a diverging linear susceptibility close to a critical point (Fig.~\ref{ss_zerofield}), qualitatively similar to the classical system. Can this be advantageous in quantum sensing applications? More precisely, can one enhance the sensitivity of a damped quantum oscillator by introducing an incoherent pump? For such a passive oscillator, the equation of motion is given by $\smash{\dot{\hat{a}} = \Omega - (\gamma_1^-/2)\hat{a}}$, so the steady state is $\langle\hat{a}\rangle=2\Omega/\gamma_1^-$ and the passive susceptibility is \smash{$\chi_p = 2/\gamma_1^-$}. To discuss whether pumping enhances sensitivity, we define the sensitivity gain at weak drives, $G_0 \equiv \chi/\chi_p|_{\Omega\to 0}$.

\begin{figure}
\centering
\includegraphics[width=\columnwidth]{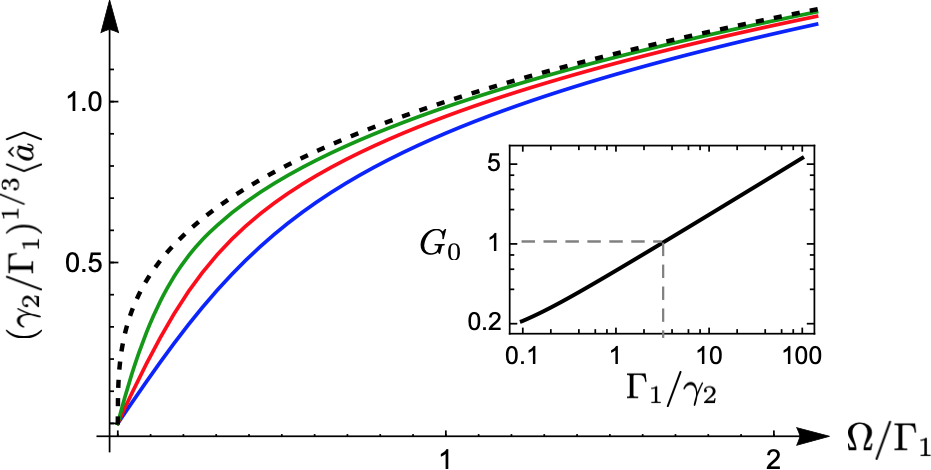}
\caption{\label{gain_critical}Response for the critical condition $\gamma_1^+ = \gamma_1^- = \Gamma_1$ with $\Gamma_1/\gamma_2=$ 5 (blue), 50 (red), and 1000 (green), showing the approach to classical limit (dotted black). Inset shows the sensitivity gain over a passive oscillator, $G_0\sim\smash{\sqrt{\Gamma_1/\gamma_2}}$.
}
\end{figure}

One regime where there is a gain, $G_0>1$, is the classical limit. Then the response is only a function of $\gamma_1^+ - \gamma_1^-$, and one can simply set $\gamma_1^+ = \gamma_1^-$ to make the system infinitely sensitive to weak signals for any damping. Quantum mechanically, however, the response also depends on the average $\Gamma_1 \equiv (\gamma_1^+ + \gamma_1^-)/2$. Figure~\ref{gain_critical} shows how the classical limit emerges with increasing $\Gamma_1$ for the critical case, $\gamma_1^+=\gamma_1^-$. We see that $G_0$ scales as $\smash{\sqrt{\Gamma_1/\gamma_2}}$ 
and one can have  $\chi>\chi_p$ if $\gamma_1^- \gtrsim 3\gamma_2$, which holds for typical experimental systems \cite{Leghtas2015}. The relatively slow growth of $G_0$ with $\Gamma_1$ stems from large number fluctuations at criticality (which also limits the sensing efficiency, see below). Since $\gamma_1^+=\gamma_1^-$, the undriven system would reach an infinite-temperature state in the limit $\gamma_2\to 0$. A nonzero $\gamma_2$ results in a Gaussian number distribution of width $\smash{\sqrt{\Gamma_1/\gamma_2}}$~\cite{dodonov1997exact}, leading to the linear susceptibility $\chi\approx 2/\sqrt{\pi\Gamma_1\gamma_2}$ (derivation in Supplement \cite{Note1}).

\begin{figure}
\centering
\includegraphics[width=\columnwidth]{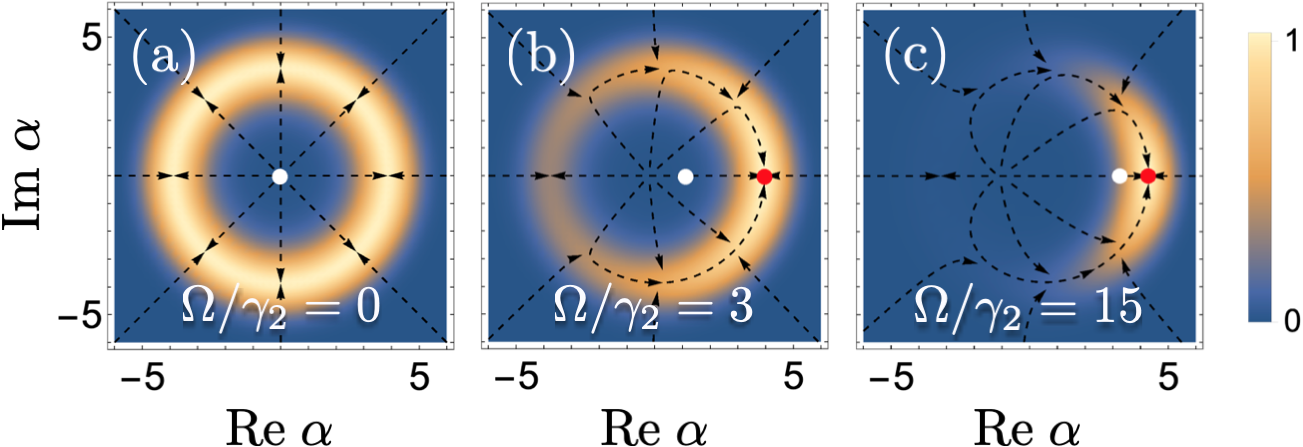}
\caption{\label{gain_limitcycle}(a)--(c) Classical trajectories and rescaled Wigner functions at increasing drive for $\gamma_1^+/\gamma_2=50$ and $\gamma_1^-/\gamma_2=20$. The quantum and classical responses are shown by white and red dots, respectively. As the quantum system crosses over from a symmetry-broken limit cycle to the classical steady state, its susceptibility is only limited by two-body loss.}
\end{figure}

The sensitivity can be enhanced much further by operating deep in the limit-cycle phase, $\gamma_1^+ - \gamma_1^- \gg \gamma_2$, as in Fig.~\ref{gain_limitcycle}. Here, a vanishingly small drive breaks the phase-rotation symmetry, yielding a nonzero classical response. For the quantum oscillator, the coherence $\langle\hat{a}\rangle$ builds up at a finite rate (see Supplement for a derivation \cite{Note1})
\begin{equation}
\chi \approx \textstyle\frac{2}{3}\hspace{0.03cm}\gamma_2^{-1} \big[1-2\gamma_1^-/(3\gamma_1^+)\big]\;.
\label{largess}
\end{equation}
Hence, the susceptibility is only limited by $\gamma_2$ and one can obtain a large enhancement. In particular, for $\gamma_1^+\gg\gamma_1^-$, $G_0\approx \gamma_1^-/(3\gamma_2)\gg 1$. In other words, a sufficiently strong incoherent pump can negate the linear damping and yield a response bounded only by two-body loss. This heightened sensitivity persists until the oscillator switches over to the classical limit at \smash{$\Omega \gtrsim (\gamma_1^+\gamma_2)^{1/2}$}, as shown in Fig.~\ref{gain_limitcycle}. Such a behavior is reminiscent of sensing protocols based on switching in optically bistable systems~\cite{Aldana2014}.

Apart from the ``signal'' $\langle\hat{a}\rangle$, the efficiency of sensors is limited by the signal-to-noise ratio (SNR) \cite{Degen2017, Lau2018}. Estimating the noise from the spread of the Wigner function, we find one can have $G_0>1$ and $\text{SNR}\gtrsim 1$ in the quantum domain, $\langle\hat{a}\rangle \sim 1$ (see Supplement~\cite{Note1}). Further, one can measure weak signals in fewer shots in the limit-cycle regime than in the classical limit of the critical oscillator. This is because particle-number fluctuations diverge in this limit, causing the SNR to vanish.

\begin{figure}
\centering
\includegraphics[width=\columnwidth]{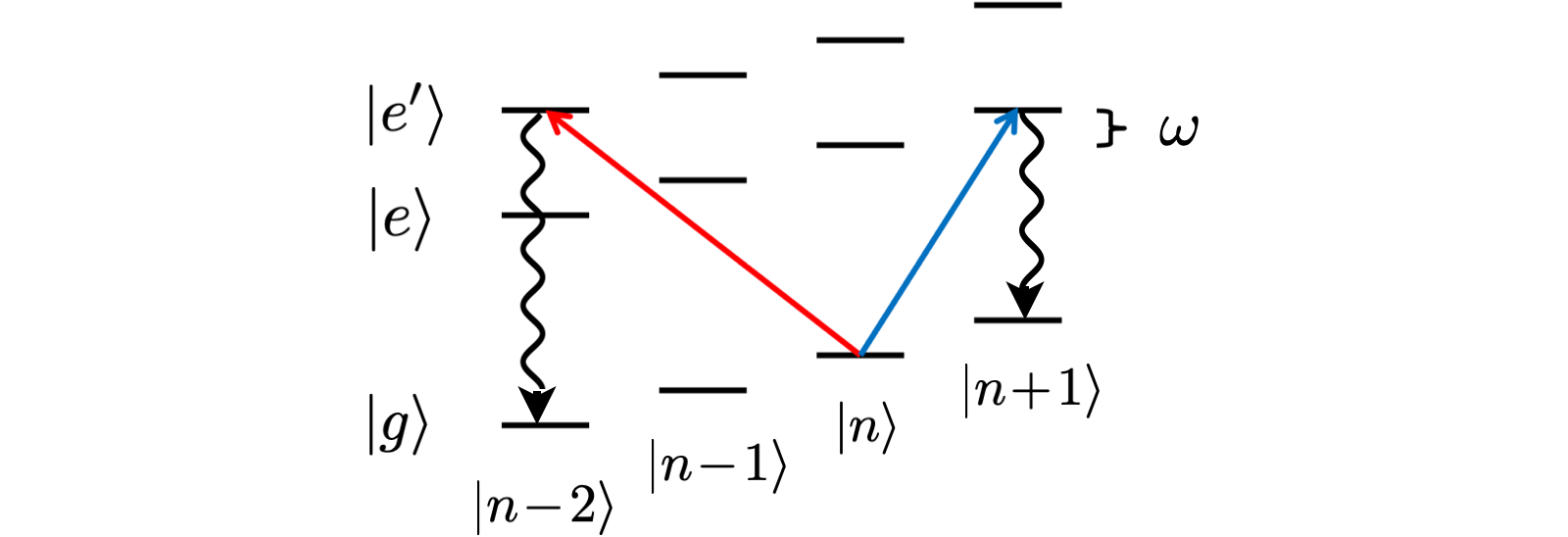}
\caption{\label{ionexp}Experimental scheme for realizing a quantum vdP oscillator with a trapped ion, adapted from Ref.~\cite{lee2013quantum}. Motional modes $|n\rangle$ act as the energy levels of the oscillator, while $|g\rangle$, $|e\rangle$, and $|e^{\prime}\rangle$ represent internal states. Sideband transitions, followed by spontaneous decay, effectively inject or remove energy quanta (phonons), realizing gain and loss processes.}
\end{figure}

The quantum vdP model can be probed in several experimental setups. In particular, as proposed in \cite{lee2013quantum, walter2014quantum}, one can engineer the dissipation via sideband transitions which either add or remove energy quanta. For instance, one can laser excite a harmonically trapped ion to its blue or red motional sidebands to implement one-phonon gain or two-phonon loss, as illustrated in Fig.~\ref{ionexp}. In such a setup, the environment relaxation time is of order $\Delta E^{-1}$, where $\Delta E$ is the level spacing between $|g\rangle$ and $|e\rangle$, which dominates all other energy scales \cite{Daley2014}. Hence, the Markov approximation holds. Additionally, one must have resolved sidebands and suppress off-resonant excitations. As shown in Ref.~\cite{lee2013quantum}, one can satisfy these constraints for several tens of low-energy modes. The response to a resonant drive can be measured with well-established techniques, including tomography \cite{Lvovsky2009} and more direct mapping of the Wigner function \cite{Banaszek1996, Banaszek1999, Leibfried1996, Lutterbach1997, Bertet2002}.

Similar techniques can be used for phonon modes of an optomechanical membrane \cite{walter2014quantum} or photon modes in microwave resonators \cite{Johansson2014}. One can also realize an incoherent pump through spontaneous emission from a set of inverted qubits \cite{Lebreuilly2016, Ma2017}. Strong two-body loss can result from Josephson junctions in superconducting circuits \cite{Leghtas2015} or polariton blockade in an optical cavity \cite{Jia2018
}.

Finally, note that Eq.~\eqref{mastereq} is not the only quantum model that reproduces the classical limit of the vdP oscillator. In particular, one can harness energy-dependent one-body loss instead of two-body loss, as in Refs.~\cite{Loerch2017, Nigg2018, Viennot2018, Rips2012}. Numerics show that the key features are unaffected. There may also exist alternative descriptions of the vdP oscillator in terms of nonstandard Hamiltonians with possible quantization \cite{Chandrasekar2007, Bender2016, Ruby2012}.
%

In summary, we have characterized the response of an open quantum system across a dynamical critical point by modeling a prototypical self-sustained oscillator. Like its classical counterpart, we find an increased sensitivity to external drives, which is ideal for sensing applications. 
Genuine quantum features arise at very weak drives, including a power-law diverging linear susceptibility and a concomitant negative susceptibility. To further explore the prospects of sensing, we compare the response with that of a passive system, where the sensitivity is limited by one-body loss. We show this linear damping is negated by a strong incoherent pump, yielding a response only bounded by weak nonlinear effects, e.g., two-body decay. Our findings are largely generic and can be probed in present-day experimental platforms, contributing 
to a broader understanding of dynamical criticality \cite{roli2018dynamical} and quantum-to-classical crossover \cite{Davidovich2016, schlosshauer2014decoherence}.

\vspace{0.2cm}
We thank Andreas Nunnenkamp and Chris Parmee for helpful discussions. This work was supported by EPSRC Grant EP/P009565/1 and a Simons Investigator Award.

\begingroup
\renewcommand{\addcontentsline}[3]{}
\renewcommand{\section}[2]{}


%

\endgroup


\onecolumngrid
\clearpage

\begin{center}
\textbf{\large Supplemental Material for\\ ``Critical response of a quantum van der Pol oscillator"}
\end{center}

\setcounter{equation}{0}
\setcounter{figure}{0}
\setcounter{table}{0}
\setcounter{page}{1}
\setcounter{secnumdepth}{3}
\makeatletter

\renewcommand{\thefigure}{S\arabic{figure}}
\renewcommand{\theequation}{S\arabic{equation}}
\renewcommand{\bibnumfmt}[1]{[S#1]}
\renewcommand{\citenumfont}[1]{S#1}
\renewcommand{\thesection}{S\Roman{section}}

\renewcommand{\theHfigure}{S\thefigure}

%
%
%
%

\tableofcontents

\section{\label{classical}Steady state of a classical van der Pol oscillator}
\begin{figure}
\includegraphics[height=0.28\columnwidth
]{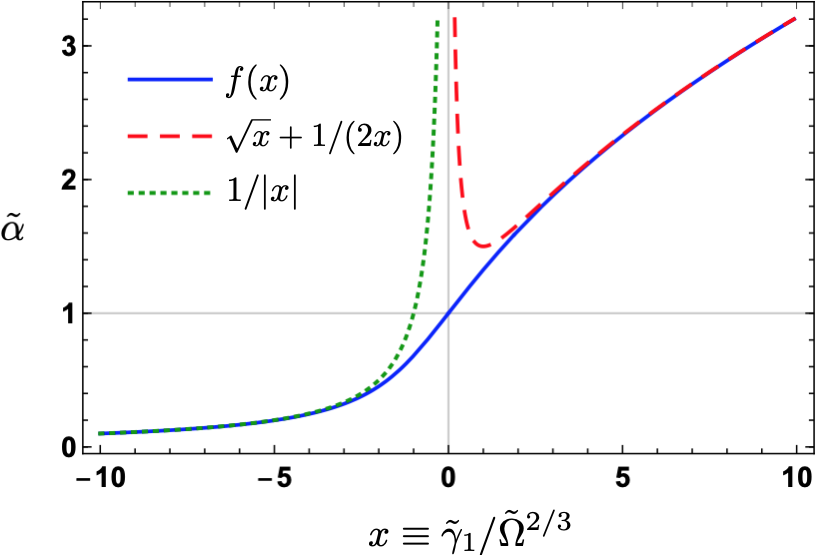}
\caption{\label{fclassicalfig}Classical steady-state response as a function of rescaled parameters. The function $f(x)$ is defined in Eq.~\eqref{fclassical}.}
\end{figure}

Here we derive a closed-form expression for the steady-state response of a classical van der Pol (vdP) oscillator. As described in the main text, the equation of motion for such an oscillator is given by
\begin{equation}
\dot{\alpha} = \gamma_1 \alpha - \gamma_2 |\alpha|^2 \alpha + \Omega \;,
\end{equation}
where $\alpha$ is the complex amplitude, $\gamma_1$ is the negative damping, $\gamma_2$ is the nonlinear damping, and $\Omega$ is the drive. The phase of the drive is chosen such that $\Omega$ is real and positive. Hence, $\alpha$ is real in steady state. To solve for the response, we introduce the rescaled quantities $\tilde{\gamma}_1 \equiv \gamma_1/\gamma_2$, $\tilde{\Omega} \equiv \Omega/\gamma_2$, and $\tilde{\alpha} \equiv \alpha / \tilde{\Omega}^{1/3}$. In steady state,
\begin{equation}
(\tilde{\gamma}_1/\tilde{\Omega}^{2/3})\hspace{0.03cm} \tilde{\alpha} - \tilde{\alpha}^3 +1 = 0 \;.
\label{classicalsteadystate}
\end{equation}
The only stable solution to Eq.~\eqref{classicalsteadystate} is given by $\tilde{\alpha} = f(\tilde{\gamma}_1/\tilde{\Omega}^{2/3})$, where
\begin{equation}
f(x) \equiv \bigg(\frac{1}{2} + \sqrt{\frac{1}{4} - \frac{x^3}{27}}\hspace{0.08cm}\bigg)^{\hspace{-0.08cm}1/3} + \hspace{0.05cm}\text{sgn}(x)^{2/3} \bigg(\frac{1}{2} - \sqrt{\frac{1}{4} - \frac{x^3}{27}}\hspace{0.08cm}\bigg)^{\hspace{-0.08cm}1/3}.
\label{fclassical}
\end{equation}
Here, sgn denotes the sign function and the powers are calculated using principal values of the arguments. Figure~\ref{fclassicalfig} shows the variation of $f(x)$. Note that $f(0)=1$, i.e., $\alpha = (\Omega/\gamma_2)^{1/3}$ for $\gamma_1=0$. This is the characteristic nonlinear response utilized in biological sensors \cite{seguiluz2000essential}. Conversely, for large $x$, $f(x)$ can be approximated as
\begin{equation}
f(x)= 
   \begin{cases} 
      \sqrt{x} + 1/(2x) + O(x^{-5/2}) & \text{for}\;\;x \gg 1\quad, \\
      1/|x| + O(x^{-4}) & \text{for}\;\;x \ll -1\;.
   \end{cases}
\end{equation}
The former limit is realized for $\gamma_1 > 0$ and $\Omega/\gamma_1 \ll \sqrt{\gamma_1/\gamma_2}$, where the steady state corresponds to a perturbed limit cycle, $\alpha = \sqrt{\gamma_1/\gamma_2} + \Omega/(2\gamma_1) + O\big[(\Omega/\gamma_1)^2 \sqrt{\gamma_2/\gamma_1}\hspace{0.03cm}\big]$. The latter limit is realized for $\gamma_1<0$ and $\Omega/|\gamma_1| \ll \sqrt{|\gamma_1|/\gamma_2}$. Here the system is fully damped and the response grows linearly with drive, $\alpha = \Omega/|\gamma_1| + O[(\Omega/\gamma_1)^3 (\gamma_2 / \gamma_1)]$.

\section{\label{quantumweak}Response of a quantum van der Pol oscillator at weak drives}
In this section, we find analytic expressions describing the response of a weakly driven quantum vdP oscillator in special cases of interest. We start from the full dynamics in rotating frame (see main text),
\begin{equation}
\dot{\hat{\rho}}= [\Omega(\hat{a}^{\dagger}-\hat{a}),\hat{\rho}] + \gamma_1^+\hspace{0.03cm}\mathcal{D}[\hat{a}^{\dagger}]\hat{\rho}  + \gamma_1^-\hspace{0.03cm}\mathcal{D}[\hat{a}]\hat{\rho} + \gamma_2\hspace{0.03cm}\mathcal{D}[\hat{a}^2]\hat{\rho}\;,
\label{fullmaster}
\end{equation}
where $\hat{\rho}$ is the density matrix, $\hat{a}$ is the ladder operator, $\Omega$ is the drive amplitude, $\gamma_1^+$ is the rate of one-particle gain, $\gamma_1^-$ is the rate of one-particle decay, $\gamma_2$ is the rate of two-particle decay, and $\mathcal{D}[\hat{x}]\hat{\rho} \equiv \hat{x}\hat{\rho}\hat{x}^{\dagger} - \{\hat{x}^{\dagger}\hat{x},\hat{\rho}\}/2$. In the Fock basis $\{|n\rangle, n=0,1,2,\dots\}$, Eq.~\eqref{fullmaster} can be written as
\begin{align}
\nonumber \dot{\rho}_{n,n^{\prime}} =&\; \Omega \left(\sqrt{n}\hspace{0.03cm} \rho_{n-1,n^{\prime}} - \sqrt{n+1} \hspace{0.03cm}\rho_{n+1,n^{\prime}} + \sqrt{n^{\prime}}\hspace{0.03cm}\rho_{n,n^{\prime}-1} - \sqrt{n^{\prime}+1}\hspace{0.03cm}\rho_{n,n^{\prime}+1}\right) \\
\nonumber & + \gamma_1^+ \Big[ \sqrt{n n^{\prime}}\hspace{0.03cm} \rho_{n-1,n^{\prime}-1} - \Big(\frac{n+n^{\prime}}{2} +1\Big) \rho_{n,n^{\prime}} \Big] \\
\nonumber & + \gamma_1^- \Big[ \sqrt{(n+1) (n^{\prime}+1)}\hspace{0.05cm} \rho_{n+1,n^{\prime}+1} - \frac{n+n^{\prime}}{2}\hspace{0.03cm} \rho_{n,n^{\prime}} \Big] \\
& + \gamma_2\hspace{0.04cm} \Big[ \sqrt{(n+2) (n+1) (n^{\prime} +2) (n^{\prime}+1)} \hspace{0.05cm} \rho_{n+2,n^{\prime}+2} - \frac{n(n-1) + n^{\prime} (n^{\prime} -1)}{2} \hspace{0.03cm} \rho_{n,n^{\prime}} \Big]\;.
\label{masterfockbasis}
\end{align}
Note that $\Omega$ couples neighboring elements of the density matrix, whereas the dissipative terms only couple elements in the same diagonal, i.e., with $n-n^{\prime} = $ constant. The response is given by $\langle\hat{a}\rangle = \text{Tr}(\hat{a}\hat{\rho}) = \sum_n \sqrt{n} \rho_{n,n-1}$, where $\rho$ is the steady-state density matrix. Since all coefficients in Eq.~\eqref{masterfockbasis} are real, $\rho$ will be real and symmetric. Below we consider this steady-state response in various limits.

\subsection{\label{simplestlinres}Linear response for $\gamma_1^+ = \gamma_1^- =0$}
For $\gamma_1^{\pm}=0$ and weak drives $\Omega \ll \gamma_2$, any two-particle excitation decays very rapidly. Hence, the dynamics are well approximated by retaining the lowest three oscillator levels, $n=0,1,2$, in Eq.~\eqref{masterfockbasis}, which yields
\begin{subequations}
\begin{align}
\label{rho00simplest}\dot{\rho}_{00} &= 2\gamma_2 \rho_{22} - 2\Omega \rho_{10}\;, \\
\label{rho11simplest}\dot{\rho}_{11} &= 2\Omega(\rho_{10} - \sqrt{2}\rho_{21})\;, \\
\label{rho10simplest}\dot{\rho}_{10} &= \Omega (\rho_{00} - \rho_{11})\;, \\
\label{rho21simplest}\dot{\rho}_{21} &= -\gamma_2\rho_{21} + \Omega (\sqrt{2} \rho_{11} + \rho_{20} - \sqrt{2} \rho_{22})\;, \\
\label{rho20simplest}\dot{\rho}_{20} & = -\gamma_2\rho_{20} + \Omega (\sqrt{2} \rho_{10} - \rho_{21})\;,
\end{align}
\end{subequations}
where $\rho_{22} = 1-\rho_{00} - \rho_{11}$. Substituting $\Omega \to 0$ into these equations, we find the undriven steady state is given by $\rho_{22} = \rho_{21} = \rho_{20} = 0$ and $\rho_{00}=\rho_{11}=1/2$. To linear order in $\Omega/\gamma_2$, Eqs.~\eqref{rho11simplest} and \eqref{rho21simplest} yield $\rho_{10} = \sqrt{2}\rho_{21} = \Omega/\gamma_2$. Thus, the linear response is given by $\langle\hat{a}\rangle = \rho_{10} + \sqrt{2}\rho_{21} = 2\Omega/\gamma_2$, as evident in Fig.~\ref{ss_critical} of the main text.

\subsection{\label{nonmonotonicres}Nonmonotonic response for $\gamma_1^{\pm} \ll \gamma_2$}
\subsubsection{\label{nonmonotonicresdamped}$\gamma_1^+=0$, $0<\gamma_1^- \ll \gamma_2$}
\begin{figure}[h]
\includegraphics[height=0.28\columnwidth
]{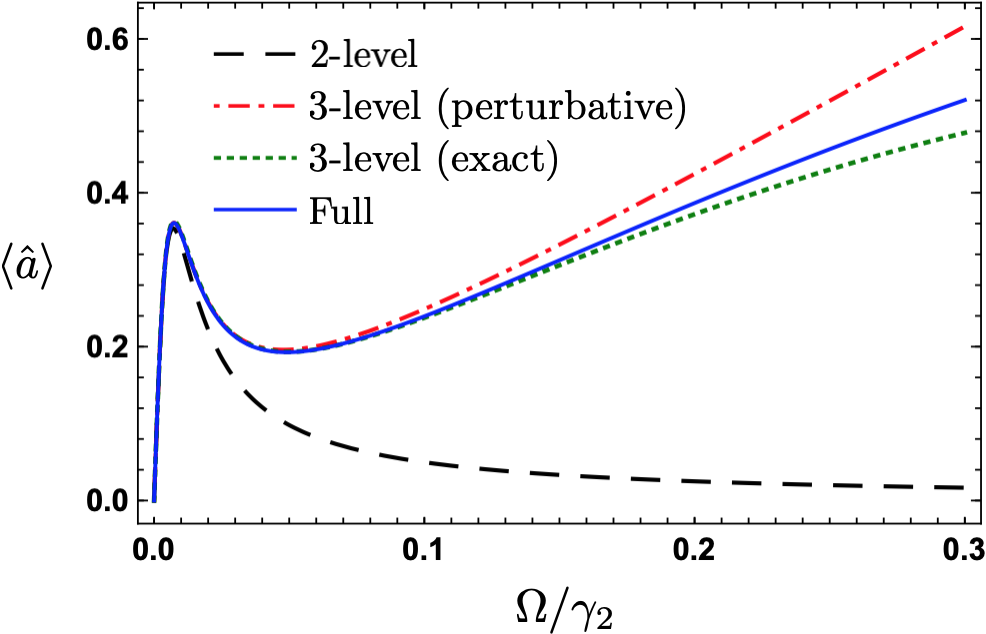}
\caption{\label{nonmonotonicdamped}Response $\langle\hat{a}\rangle$ as a function of drive $\Omega$ for $\gamma_1^+=0$ and $\gamma_1^-/\gamma_2=0.02$. The perturbative result corresponds to Eq.~\eqref{responsethreeleveldamped}.
}
\end{figure}

{\it Two-level approximation.\textemdash} As the system is fully damped, we can project the dynamics at weak drives onto the manifold spanned by $n=$ 0 and 1. Then Eq.~\eqref{masterfockbasis} reduces to
\begin{subequations}
\begin{align}
\label{rho00twolevel}\dot{\rho}_{11} &= 2\Omega \rho_{10} - \gamma_1^- \rho_{11}\;, \\
\label{rho10twolevel}\dot{\rho}_{10} &= \Omega (\rho_{00} - \rho_{11}) - (\gamma_1^-/2)\rho_{10}\;,
\end{align}
\end{subequations}
where $\rho_{00}=1-\rho_{11}$. In steady state, we find
\begin{equation}
\label{responsetwolevel}\langle\hat{a}\rangle = \rho_{10} = \frac{2\Omega\gamma_1^-}{(\gamma_1^-)^2 + 8 \Omega^2}\;.
\end{equation}
Thus, the response grows linearly as $\langle\hat{a}\rangle \approx 2\Omega/\gamma_1^-$ for $\Omega \lesssim \gamma_1^-$, but falls off as $\langle\hat{a}\rangle \approx\gamma_1^-/(4\Omega)$ for $\Omega \gtrsim \gamma_1^-$. This is shown by the dashed black curve in Fig.~\ref{nonmonotonicdamped}. We see the linear susceptibility diverges as $2/\gamma_1^-$ for $\gamma_1^- \to 0$.

{\it Three-level approximation.\textemdash} The two-level picture breaks down at large drives. As higher-energy modes come into play, the response grows again with $\Omega$, as seen in Fig.~\ref{nonmonotonicdamped}. This revival can be captured by including the $n=2$ level in the dynamics, yielding, with $\rho_{11} = 1 - \rho_{00} - \rho_{22}$,
\begin{subequations}
\begin{align}
\label{rho00threeleveldamped}\dot{\rho}_{00} &= 2\gamma_2 \rho_{22} - 2\Omega \rho_{10} + \gamma_1^- \rho_{11}\;, \\
\label{rho10threeleveldamped}\dot{\rho}_{10} &= \Omega (\rho_{00} - \rho_{11}) + \gamma_1^- (\sqrt{2}\rho_{21} - \rho_{10}/2)\;, \\
\label{rho22threeleveldamped}\dot{\rho}_{22} &= -2(\gamma_2+\gamma_1^-)\rho_{22} + 2\sqrt{2}\Omega \rho_{21}\;, \\
\label{rho21threeleveldamped}\dot{\rho}_{21} &= -(\gamma_2+3\gamma_1^-/2)\rho_{21} + \Omega \big[\sqrt{2} (\rho_{11}-\rho_{22}) + \rho_{20}\big]\;, \\
\label{rho20threeleveldamped}\dot{\rho}_{20} & = -(\gamma_2+\gamma_1^-)\rho_{20} + \Omega (\sqrt{2} \rho_{10} - \rho_{21})\;.
\end{align}
\end{subequations}
Let us solve for the steady state to lowest order in $\gamma_1^-/\gamma_2$ and $\Omega/\gamma_2$. From the last three equations, we can write
\begin{subequations}
\begin{align}
\label{rho22threeleveldamped1}\rho_{22} &\approx (\sqrt{2} \Omega/\gamma_2) \rho_{21}\;,\\
\label{rho21threeleveldamped1}\rho_{21} &\approx (\Omega/\gamma_2) \big[\sqrt{2} (\rho_{11}-\rho_{22}) + \rho_{20}\big]\;, \\
\label{rho20threeleveldamped1}\rho_{20} &\approx (\Omega/\gamma_2) (\sqrt{2} \rho_{10} - \rho_{21})\;.
\end{align}
\end{subequations}
Substituting $\rho_{22}$ and $\rho_{20}$ into Eq.~\eqref{rho21threeleveldamped1}, we find
\begin{equation}
\label{rho21threeleveldamped2}\rho_{21} \approx (\sqrt{2}\Omega/\gamma_2) [\rho_{11} + (\Omega/\gamma_2) \rho_{10}]\;.
\end{equation}
Using Eqs.~\eqref{rho22threeleveldamped1} and \eqref{rho21threeleveldamped2} in Eq.~\eqref{rho00threeleveldamped} yields
\begin{equation}
\label{rho10threeleveldamped3}\rho_{10} \approx \left(\frac{2\Omega}{\gamma_2} + \frac{\gamma_1^-}{2\Omega}\right) \rho_{11}\;.
\end{equation}
Substituting this expression back into Eqs.~\eqref{rho21threeleveldamped2} and \eqref{rho22threeleveldamped1} yields
\begin{align}
\label{rho21threeleveldamped4}\rho_{21} &\approx (\sqrt{2} \Omega/\gamma_2) \rho_{11}\;,\\
\label{rho22threeleveldamped4}\rho_{22} &\approx 2(\Omega/\gamma_2)^2 \rho_{11}\;.
\end{align}
Finally, substituting Eqs.~\eqref{rho10threeleveldamped3}--\eqref{rho22threeleveldamped4} into Eq.~\eqref{rho10threeleveldamped} and using $\rho_{00} = 1-\rho_{11} - \rho_{22}$, we find
\begin{equation}
\label{rho11threeleveldamped5}\rho_{11} \approx \frac{4\Omega^2}{(\gamma_1^-)^2 + 8 \Omega^2}\;,
\end{equation}
which matches the two-level result. However, the response is modified as
\begin{equation}
\label{responsethreeleveldamped}\langle\hat{a}\rangle = \rho_{10} + \sqrt{2} \rho_{21} \approx \left(\frac{4\Omega}{\gamma_2} + \frac{\gamma_1^-}{2\Omega}\right) \rho_{11} \approx \frac{2\Omega}{\gamma_2} \left[\frac{\gamma_1^- \gamma_2 + 8 \Omega^2}{(\gamma_1^-)^2 + 8 \Omega^2}\right],
\end{equation}
where we have used Eqs.~\eqref{rho10threeleveldamped3}, \eqref{rho21threeleveldamped4}, and \eqref{rho11threeleveldamped5}. It reduces to the two-level expression in Eq.~\eqref{responsetwolevel} for $\Omega \ll (\gamma_1^- \gamma_2)^{1/2}$. For $\Omega \sim (\gamma_1^- \gamma_2)^{1/2} \gg \gamma_1$, we find $\langle\hat{a}\rangle \approx \gamma_1^-/(4\Omega) + 2\Omega/\gamma_2$, which attains a minimum at $\Omega = (\gamma_1^- \gamma_2)^{1/2}/(2\sqrt{2})$ and increases linearly with drive at large $\Omega$. This nonmonotonic variation is evident in Fig.~\ref{nonmonotonicdamped}.

\subsubsection{\label{nonmonotonicresfull}$0\leq \gamma_1^+,\gamma_1^- \ll \gamma_2$}
In the presence of one-particle gain ($\gamma_1^+>0$), the undriven steady state corresponds to a dynamic equilibrium where particles flow in and out of the oscillator. When $\gamma_2 \gg \gamma_1^{\pm}$, the dynamics are confined to the levels $n=$ 0, 1, and 2 for weak drives. We consider this three-level system, governed by the equations of motion [from Eq.~\eqref{masterfockbasis}]
\begin{subequations}
\begin{align}
\label{rho00threelevel}\dot{\rho}_{00} &= 2\gamma_2 \rho_{22} - 2\Omega \rho_{10} + \gamma_1^- \rho_{11} - \gamma_1^+ \rho_{00}\;, \\
\label{rho10threelevel}\dot{\rho}_{10} &= \Omega (\rho_{00} - \rho_{11}) + \gamma_1^- (\sqrt{2}\rho_{21} - \rho_{10}/2) - (3\gamma_1^+/2)\rho_{10}\;, \\
\label{rho22threelevel}\dot{\rho}_{22} &= -2(\gamma_2+\gamma_1^-)\rho_{22} + 2\sqrt{2}\Omega \rho_{21} + 2\gamma_1^+ \rho_{11}\;, \\
\label{rho21threelevel}\dot{\rho}_{21} &= -(\gamma_2+3\gamma_1^-/2)\rho_{21} + \Omega \big[\sqrt{2} (\rho_{11}-\rho_{22}) + \rho_{20}\big] + \sqrt{2}\gamma_1^+ \rho_{10}\;, \\
\label{rho20threelevel}\dot{\rho}_{20} & = -(\gamma_2+\gamma_1^-)\rho_{20} + \Omega (\sqrt{2} \rho_{10} - \rho_{21})\;,
\end{align}
\end{subequations}
where $\rho_{11} = 1 - \rho_{00} - \rho_{22}$. For $\Omega=0$, we find the steady state $\rho_{11} \approx 1-\rho_{00} \approx \gamma_1^+/(3\gamma_1^+ + \gamma_1^-)$. Thus, negative damping leads to a finite occupation of the $n=1$ level. For $\Omega >0$, the steady-state analysis can be carried out following the same steps as in the last section, which yields, to lowest order in $\gamma_1^+/\gamma_2$, $\gamma_1^-/\gamma_2$, and $\Omega/\gamma_2$,
\begin{equation}
\label{responsethreelevel}\langle\hat{a}\rangle \approx \frac{2\Omega}{\gamma_2} \left[\frac{(\gamma_1^+ + \gamma_1^-) \gamma_2 + 8 \Omega^2}{(3\gamma_1^+ + \gamma_1^-)^2 + 8 \Omega^2}\right].
\end{equation}
As before, we find a nonmonotonic response. For $\Omega \ll 3\gamma_1^+ + \gamma_1^-$, the response is linear, $\langle\hat{a}\rangle \approx  2\Omega (\gamma_1^+ + \gamma_1^-) / (3\gamma_1^+ + \gamma_1^-)^2$, with a slope that diverges as $\gamma_1^{\pm} \to 0$. For $3\gamma_1^+ + \gamma_1^- \ll \Omega \ll [(\gamma_1^+ + \gamma_1^-) \gamma_2]^{1/2}$, it falls off as $\langle\hat{a}\rangle \approx (\gamma_1^+ + \gamma_1^-)/(4\Omega)$, exhibiting negative susceptibility. For $\Omega \gg [(\gamma_1^+ + \gamma_1^-) \gamma_2]^{1/2}$, it rises again as $\langle\hat{a}\rangle \approx 2\Omega/\gamma_2$.

\subsection{\label{criticalsensing}Linear response for $\gamma_1^+ = \gamma_1^- \gg \gamma_2$}
\begin{figure}[b]
\includegraphics[height=0.28\columnwidth]{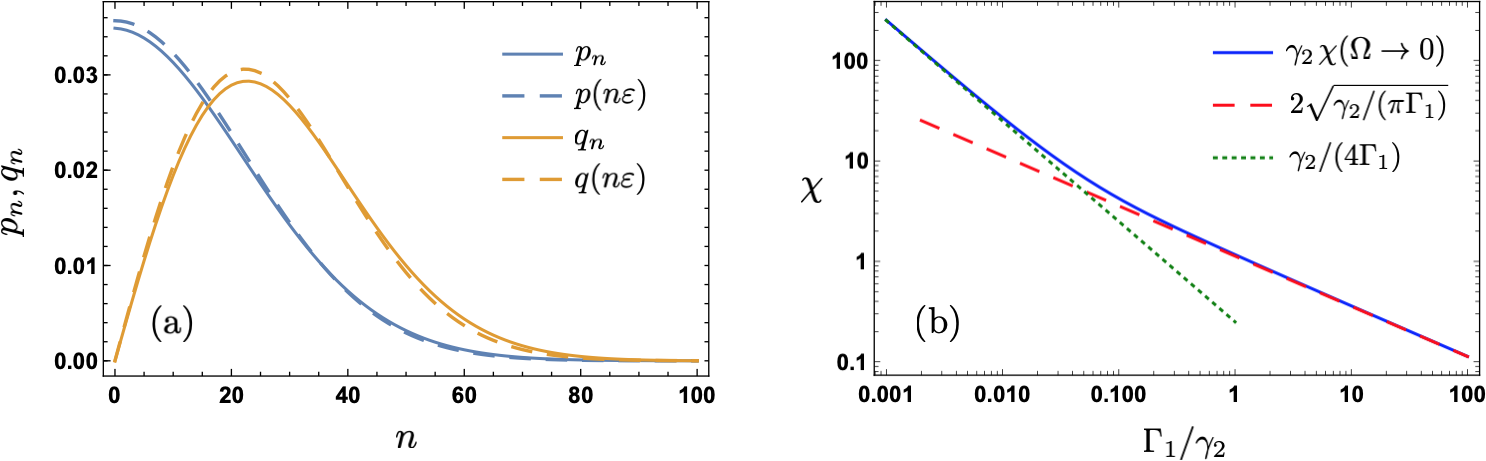}
\caption{\label{critfig} (a) Steady-state occupations $p_n \equiv \rho_{n,n}$ and coherences $q_n \equiv \sqrt{n} \rho_{n,n-1}$ for $\Gamma_1/\gamma_2 = 1000$ and $\Omega\to 0$. The coherences are rescaled to show on the same axes. Solid lines are obtained from exact numerics and dashed lines correspond to Eqs.~\eqref{gaussian_crit} and \eqref{qx}. (b) Linear susceptibility $\chi$ as a function of $\Gamma_1/\gamma_2$. Solid curve shows exact result and dashed curves show aymptotes.}
\end{figure}

Here, we derive the enhanced sensitivity for the critical condition $\gamma_1^+ = \gamma_1^- \equiv \Gamma_1 \gg \gamma_2$. We first find the undriven steady state, then consider perturbation at weak drives. At $\Omega = 0$, the steady state has no coherence. The equation of motion for the populations $p_n \equiv \rho_{n,n}$ is found by substituting $n^{\prime} = n$ in Eq.~\eqref{masterfockbasis}, which yields
\begin{equation}
\dot{p}_n = \Gamma_1 \big[n p_{n-1} + (n+1) p_{n+1} - (2 n + 1) p_n\big] + \gamma_2 \big[(n+1)(n+2)p_{n+2} - n(n-1)p_n\big]\;.
\label{eompopulations_crit}
\end{equation}
At $\gamma_2=0$, the steady state is simply given by $p_n = $ constant, i.e., the system reaches an infinite-temperature state. For $\gamma_2>0$, this distribution is curtailed at large $n$ and $p_n$ is a slowly-varying function set by the scale $\varepsilon \equiv \sqrt{\gamma_2/\Gamma_1}$. Thus, we can approximate $p_n$ by a continuous function $p(x)$ with $x=n\varepsilon$. Then Eq.~\eqref{eompopulations_crit} can be written as
\begin{equation}
x p(x-\varepsilon) + (x+\varepsilon) p(x+\varepsilon) - (2x+\varepsilon) p(x) + \varepsilon \big[(x+\varepsilon) (x+2\varepsilon) p(x+2\varepsilon) - x (x-\varepsilon) p(x) \big] = 0\;.
\label{eomcontpop_crit}
\end{equation}
The above equation is identically satisfied up to $O(\varepsilon)$. Expanding to $O(\varepsilon^2)$, we find
\begin{equation}
x p''(x) + (2 x^2 + 1) p'(x) + 4 x p(x) = 0\;,
\end{equation}
where prime denotes d/d$x$. Requiring $p(x)$ to be bounded at $x=0$ leads to the unique solution $p(x) = C e^{-x^2}$. The constant $C$ can be determined from probability conservation $\sum_n p_n =1$, or $(1/\varepsilon) \int_0^{\infty} p(x) dx = 1$, which gives
\begin{equation}
p(x) = (2\varepsilon/\sqrt{\pi}\hspace{0.02cm})\hspace{0.03cm} e^{-x^2}.
\label{gaussian_crit}
\end{equation}
Thus, the undriven steady state has a Gaussian number distribution of width $\sqrt{\Gamma_1/\gamma_2} \gg 1$, as shown in Fig.~\ref{critfig}(a).

To find the linear response, we consider perturbation in the leading off-diagonal terms $\chi_n \equiv\rho_{n,n-1}$, which satisfy
\begin{equation}
\dot{\chi}_n = \Gamma_1 \big[\sqrt{n(n-1)} \chi_{n-1} + \sqrt{n(n+1)} \chi_{n+1} - 2 n \chi_n\big] + \gamma_2 \big[\sqrt{n(n+2)}(n+1)\chi_{n+2} - (n-1)^2 \chi_n\big] + \Omega\sqrt{n} (p_{n-1}-p_n)\;.
\end{equation}
This equation is simplified with the substitution $q_n \equiv \sqrt{n} \chi_n$. In steady state,
\begin{equation}
\Gamma_1 (q_{n+1} + q_{n-1} - 2 q_n) + \gamma_2 \big[(n+1) q_{n+2} -(n+1/n-2) q_n\big] + \Omega (p_{n-1} - p_n) =0\;.
\label{eomqn_crit}
\end{equation}
Since $q_n$ will also be a slowly-varying function as $p_n$, we rewrite Eq.~\eqref{eomqn_crit} in terms of the continuous variable $x=n\varepsilon$,
\begin{equation}
q(x+\varepsilon) + q(x-\varepsilon) - 2q(x) + \varepsilon\big[(x+\varepsilon) q(x+2\varepsilon) - (x - 2\varepsilon + \varepsilon^2/x) q(x) \big] + \eta \varepsilon \big[\tilde{p}(x-\varepsilon) - \tilde{p}(x)\big] = 0\;,
\label{eomcontqn_crit}
\end{equation}
where $\eta \equiv \Omega/\Gamma_1$ and $\tilde{p}(x) \equiv p(x)/\varepsilon \sim O(1)$. Expanding Eq.~\eqref{eomcontqn_crit} to $O(\varepsilon^2)$, we find
\begin{equation}
q''(x) + 2 x q'(x) + 3 q(x) - \eta\hspace{0.05cm}\tilde{p}\hspace{0.02cm}'(x)=0\;.
\end{equation}
Substituting $\tilde{p}\hspace{0.02cm}'(x)=-(4/\sqrt{\pi}) x e^{-x^2}$ from Eq.~\eqref{gaussian_crit} and using $q(x) \equiv (4\eta/\sqrt{\pi}) e^{-x^2} h(x)$ yields
\begin{equation}
h''(x) - 2 x h'(x) + h(x) + x = 0\;.
\label{heqn}
\end{equation}
In addition, since $q(0)=0$, we have the boundary condition $h(0)=0$. Then the only bounded solution to Eq.~\eqref{heqn} is $h(x)=x$. Therefore,
\begin{equation}
q(x) = (4\eta/\sqrt{\pi}) x e^{-x^2}\;,
\label{qx}
\end{equation}
which agrees well with numerics, as shown in Fig.~\ref{critfig}(a). The linear response is given by
\begin{equation}
\langle\hat{a}\rangle = \sum_{n=0}^{\infty} \sqrt{n} \chi_n \approx \frac{1}{\varepsilon} \int_0^{\infty} \hspace{-0.1cm}dx \hspace{0.03cm}q(x) = \frac{2\eta}{\sqrt{\pi} \varepsilon} = \frac{2\Omega}{\sqrt{\pi \Gamma_1 \gamma_2}}\;.
\label{linres_crit}
\end{equation}
Hence, the linear susceptibility is $\chi \approx 2/\sqrt{\pi \Gamma_1 \gamma_2}$. Figure~\ref{critfig}(b) shows how the exact solution interpolates between this regime and that of $\Gamma_1\ll \gamma_2$, where $\chi \approx 1/(4\Gamma_1)$ [see Sec.~\ref{nonmonotonicresfull}]. In contrast, a fully damped ``passive" oscillator has a susceptibility $\chi_p = 2/\gamma_1^-$ [see Sec.~\ref{nonmonotonicresdamped}]. Thus, operating the vdP oscillator in the regime $\gamma_1^+ = \gamma_1^- \gg\gamma_2$ yields a sensitivity enhancement $\chi/\chi_p \approx [\gamma_1^-/(\pi \gamma_2)]^{1/2}$.

\subsection{\label{limitcyclesensing}Linear response for $\gamma_1^+ - \gamma_1^- \gg \gamma_2$}
\begin{figure}[b]
\includegraphics[height=0.28\columnwidth]{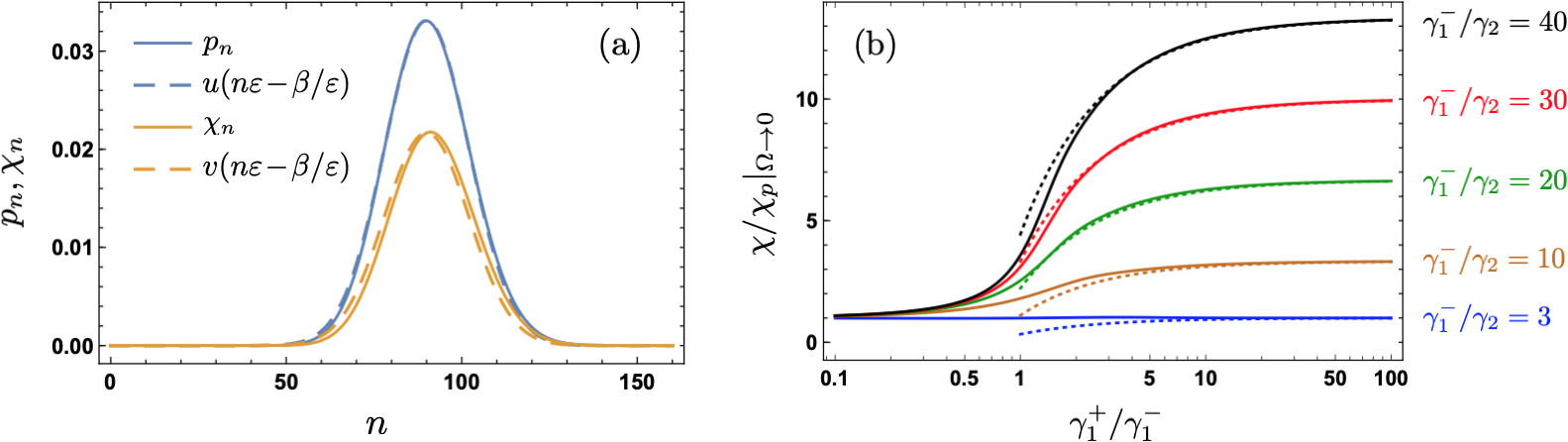}
\caption{\label{lcfigsupp}(a) Steady-state populations $p_n \equiv \rho_{n,n}$ and coherences $\chi_n \equiv \rho_{n,n-1}$ for $\gamma_1^+/\gamma_2 = 200$, $\gamma_1^-/\gamma_2=20$, and $\Omega\to 0$. The coherences have been rescaled to show on the same axes as $p_n$. Solid lines correspond to exact numerics and dashed lines show perturbative results in Eqs.~\eqref{usol} and \eqref{vsol}. (b) Linear susceptibility $\chi$ as a function of $\gamma_1^+/\gamma_1^-$ and $\gamma_1^-/\gamma_2$, rescaled by the susceptibility of a passive system, $\chi_p = 2/\gamma_1^-$. Solid lines show exact results and dotted lines are obtained from Eq.~\eqref{ss_lc}.}
\end{figure}
As we discussed in the main text, the susceptibility is enhanced much further by operating in the limit-cycle phase. Here we derive an expression for the linear response in this regime. As in the last section, we start by writing down rate equations which govern the populations of the undriven system,
\begin{equation}
\dot{p}_n = \gamma_1^+ \big[n p_{n-1} - (n+1) p_{n}\big] +\gamma_1^- \big[(n+1) p_{n+1} - n p_n\big] + \gamma_2 \big[(n+1)(n+2)p_{n+2} - n(n-1)p_n\big]\;.
\label{eompopulations_lc}
\end{equation}
The steady state is set by the parameters $\varepsilon \equiv (\gamma_2/\gamma_1^+)^{1/2} \ll 1$ and $\zeta \equiv \gamma_1^-/\gamma_1^+ < 1$. We expect, in steady state, $p_n$ will be peaked near the classical solution, $n^* \approx \alpha_{\text{cl}}^2 = (1-\zeta)/(2\varepsilon^2)$. To find this distribution, we approximate $p_n$ by a continuous function $p(x)$ where $x=n\varepsilon$. Then Eq.~\eqref{eompopulations_lc} yields
\begin{equation}
x p(x-\varepsilon) - (x+\varepsilon) p(x) +\zeta\big[(x+\varepsilon) p(x+\varepsilon) - x p(x)\big] + \varepsilon \big[(x+\varepsilon) (x+2\varepsilon) p(x+2\varepsilon) - x (x-\varepsilon) p(x) \big] = 0\;.
\label{eomcontpop_lc}
\end{equation}
Since $p(x)$ will be peaked near $x^* = n^* \varepsilon \sim O(1/\varepsilon)$, we shift the origin with the transformation $x = \beta/\varepsilon + y$, where $\beta \sim O(1)$. We also define $u(y) \equiv p(\beta/\varepsilon + y)$. Then expanding Eq.~\eqref{eomcontpop_lc} in powers of $\varepsilon$, we find
\begin{equation}
\varepsilon\beta (2\beta+\zeta-1) u'(y) + \varepsilon^2 \big[\beta(4\beta+\zeta+1)u''(y)/2 + (4\beta+\zeta-1)\{yu'(y)+u(y)\}\big] + O(\varepsilon^3) = 0\;.
\label{upowerseries}
\end{equation}
The linear term implies $\beta = (1-\zeta)/2$, exactly matching the classical estimate. Substituting this value for $\beta$ into the quadratic term leads to the differential equation
\begin{equation}
(3-\zeta) u''(y) + 4 y u'(y) + 4 u(y) = 0\;,
\label{udiff}
\end{equation}
which has a unique positive-definite solution, $u(y) = A e^{-2y^2/(3-\zeta)}$. The integration constant $A$ can be determined by requiring $\sum_n p_n = 1$, or $(1/\varepsilon) \int u(y) dy \approx 1$, which yields
\begin{equation}
u(y) \approx \varepsilon \sqrt{2/[(3-\zeta)\pi]}\hspace{0.05cm} e^{-2 y^2/(3-\zeta)}\;.
\label{usol}
\end{equation}
From the correspondence $p_n \approx u(n\varepsilon - \beta/\varepsilon)$ and the definitions $\zeta\equiv\gamma_1^-/\gamma_1^+$, $\varepsilon \equiv (\gamma_2/\gamma_1^+)^{1/2}$, we find the undriven steady state has a Gaussian number distribution with mean $(\gamma_1^+ - \gamma_1^-)/(2\gamma_2)$ and standard deviation $[(3\gamma_1^+ - \gamma_1^-)/(4\gamma_2)]^{1/2}$. This result was also found in Ref. \cite{sdodonov1997exact} and agrees well with numerics, as shown in Fig.~\ref{lcfigsupp}(a).

Next, we consider linear perturbation at weak drives. From Eq.~\eqref{masterfockbasis}, the coherences $\chi_n \equiv \rho_{n,n-1}$ satisfy
\begin{align}
\nonumber \dot{\chi}_n = &\;\gamma_1^+ \big[\sqrt{n(n-1)}\chi_{n-1} - (n+1/2)\chi_n\big] + \gamma_1^- \big[\sqrt{n(n+1)}\chi_{n+1} - (n-1/2)\chi_n\big] \\
& + \gamma_2 \big[\sqrt{n(n+2)}(n+1)\chi_{n+2} - (n-1)^2 \chi_n\big] + \Omega \sqrt{n}(p_{n-1} - p_n) \;,
\label{eomchi_lc}
\end{align}
where $p_n$ are the undriven occupations. To find the steady state, we again approximate $\chi_n$ by a continuous function $v(y)$ where $\chi_n \approx v(n\varepsilon - \beta/\varepsilon)$. Following the same steps as before, we arrive at the solution
\begin{equation}
v(y) = 4\eta (3-\zeta)^{-3/2} \sqrt{(1-\zeta)/\pi} \hspace{0.07cm} e^{-2 y^2/(3-\zeta)}\;,
\label{vsol}
\end{equation}
where $\eta \equiv \Omega/\gamma_1^+$. Thus, $\chi_n$ has the same Gaussian profile as $p_n$, as seen in Fig.~\ref{lcfigsupp}(a). The linear response is given by
\begin{equation}
\langle\hat{a}\rangle = \sum_n \sqrt{n} \chi_n \approx \frac{1}{\varepsilon^2} \int_{-\beta/\varepsilon}^{\infty} \hspace{-0.08cm}\sqrt{\beta+\varepsilon y}\hspace{0.05cm} v(y) d y \approx \frac{2\eta}{3\varepsilon^2} \left[1-\frac{2\zeta}{3} + O(\zeta^2)\right]\;.
\label{response_lc}
\end{equation}
Expressing $\eta$, $\varepsilon$, and $\zeta$ in terms of the drive and dissipation rates, we find a susceptibility
\begin{equation}
\chi \approx \frac{2}{3\gamma_2} \left(1-\frac{2\gamma_1^-}{3\gamma_1^+}\right)\;.
\label{ss_lc}
\end{equation}
Thus, for $\gamma_1^- \ll \gamma_1^+$, $\chi$ is only limited by two-particle decay. Consequently, the sensitivity gain over a passive system, $\chi/\chi_p$, scales as $\gamma_1^-/\gamma_2$ [Recall: $\chi_p = 2/\gamma_1^-$]. This scaling is illustrated in Fig.~\ref{lcfigsupp}(b).

\section{\label{wignersection}Mapping between density matrix and Wigner function}
In this section, we review the Wigner representation of the density operator and derive an explicit mapping between the Wigner function and density matrix for an oscillator mode. The Wigner distribution was introduced in Ref. \cite{sWigner1932} to enable the calculation of expectation values as integrals over a phase space, similar to classical ensemble averages. The Wigner function acts as a weight for such an integral representation of the density operator. For a single oscillator mode $\hat{a}$, it can be written as the expectation of a displaced parity operator \cite{sCahill1969}
\begin{equation}
\hat{\Pi}(\alpha) \equiv \hat{D}(\alpha) (-1)^{\hat{a}^{\dagger} \hat{a}} \hat{D}^{-1}(\alpha)\;,
\label{parityop}
\end{equation}
where $\alpha$ is a point in phase space and $\hat{D}(\alpha) \equiv e^{\alpha \hat{a}^{\dagger} - \alpha^* \hat{a}}$ is a displacement operator. The Wigner function is given by
\begin{equation}
W(\alpha) = (2/\pi)\text{Tr}\big[\hat{\rho}\hspace{0.03cm}\hat{\Pi}(\alpha)\big]\;.
\label{wigner}
\end{equation}
Conversely, the density operator $\hat{\rho}$ can be obtained from the Wigner function as
\begin{equation}
\hat{\rho} = 2 \int\hspace{-0.05cm} d^2 \alpha\hspace{0.05cm} W(\alpha)\hspace{0.05cm} \hat{\Pi}(\alpha)\;.
\label{inversewigner}
\end{equation}
So, there is a one-to-one correspondence between $\hat{\rho}$ and $W$. Note the operators $\hat{\Pi}(\alpha)$ are Hermitian with eigenvalues $\pm 1$. They can also be shown to constitute a basis for expanding any operator over the complex phase space \cite{scahill1969ordered}. In particular, they satisfy the orthonormality and completeness relations
\begin{align*}
(4/\pi) \hspace{0.03cm}\text{Tr} \big[\hat{\Pi}(\alpha),\hat{\Pi}(\alpha^{\prime})\big] &= \delta(\alpha - \alpha^{\prime}) \hat{\mathds{1}}\;,\\
(4/\pi)\hspace{-0.05cm} \int\hspace{-0.05cm} d^2 \alpha\hspace{0.03cm} \langle m | \hat{\Pi}(\alpha) | m^{\prime}\rangle \langle n |\hat{\Pi}(\alpha) | n^{\prime} \rangle^* &= \delta_{m,n} \delta_{m^{\prime},n^{\prime}}\;,
\end{align*}
where $|n\rangle$ are Fock states. Since $\hat{\Pi}(\alpha)$ are parity operators, from Eq.~\eqref{wigner} we see that $W$ is real valued and uniformly bounded, $W(\alpha) \in [-2/\pi,2/\pi]$. Further, using the property \smash{$\text{Tr}[\hat{\Pi}(\alpha)] = 1/2$} in Eq.~\eqref{inversewigner}, one finds
\begin{equation}
\int \hspace{-0.05cm}d^2 \alpha \hspace{0.05cm} W(\alpha) = 1\;.
\label{wignernorm}
\end{equation}
Hence, $W(\alpha)$ can be interpreted as a quasiprobability distribution in phase space. In particular, one can compute the expectation of any symmetrically ordered product $\{(\hat{a}^{\dagger})^n\hat{a}^m\}_S$ as an ensemble average,
\begin{equation}
\langle \{(\hat{a}^{\dagger})^n\hat{a}^m\}_S \rangle = \int \hspace{-0.05cm} d^2 \alpha\hspace{0.03cm} (\alpha^*)^n \alpha^m W(\alpha)\;,
\label{ensembleavg}
\end{equation}
which follows from using the relation $\hat{D}^{-1}(\alpha) \hat{a} \hat{D}(\alpha) = \hat{a} + \alpha \hat{\mathds{1}}$ in Eq.~\eqref{inversewigner}. The linear response $\langle\hat{a\rangle}$ corresponds to a special case of Eq.~\eqref{ensembleavg} that amounts to measuring the center of mass of the Wigner distribution,
\begin{equation}
\langle\hat{a}\rangle = \int \hspace{-0.05cm} d^2 \alpha\hspace{0.03cm} \alpha \hspace{0.03cm}W(\alpha)\;.
\label{wignercom}
\end{equation}

In principle, Eqs.~\eqref{wigner} and \eqref{inversewigner} completely specify $W(\alpha)$ in terms of $\hat{\rho}$ and {\it vice versa}. However, one can also find an explicit mapping between the Wigner function and the density matrix in the Fock basis. To see this, note the parity operator \smash{$(-1)^{\hat{a}^{\dagger}\hat{a}}$} reflects $\hat{a}$ to $-\hat{a}$, \smash{$(-1)^{\hat{a}^{\dagger}\hat{a}} \hat{a} (-1)^{\hat{a}^{\dagger}\hat{a}} = -\hat{a}$}. It follows we have the identity \cite{sCahill1969}
\begin{equation}
(-1)^{\hat{a}^{\dagger}\hat{a}} \hat{D}(\alpha) (-1)^{\hat{a}^{\dagger}\hat{a}} = \hat{D}(-\alpha) = \hat{D}^{-1}(\alpha)\;.
\label{reflection}
\end{equation}
Using the above in Eq.~\eqref{parityop}, we can write \smash{$\hat{\Pi}(\alpha) = \hat{D}(2\alpha) (-1)^{\hat{a}^{\dagger}\hat{a}}$} which, when used in Eq.~\eqref{wigner}, yields
\begin{equation}
W(\alpha) = (2/\pi) \textstyle\sum_{n,n^{\prime}} \rho_{n,n^{\prime}} \langle n^{\prime} | \hat{D}(2\alpha) | n \rangle  (-1)^{n}\;.
\end{equation}
The matrix elements of \smash{$\hat{D}$} can be expressed in terms of associated Laguerre polynomials $L^{(p)}_q(x)$ \cite{scahill1969ordered}. Thus, we find
\begin{equation}
W(\alpha) = (2/\pi)\hspace{0.03cm} e^{-2|\alpha|^2} \sum\nolimits_{n,n^{\prime}} (-1)^n \sqrt{n!/n^{\prime}!} \hspace{0.1cm}(2\alpha)^{n^{\prime}\hspace{-0.03cm}-n}\hspace{0.06cm} L_n^{(n^{\prime}\hspace{-0.03cm}-n)}(4|\alpha|^2) \hspace{0.06cm}\rho_{n,n^{\prime}}\;.
\label{wignermap}
\end{equation}
The mapping can be written in a more symmetric form in polar coordinates $\alpha=r e^{i\phi}$,
\begin{equation}
W(r,\phi) = \frac{2}{\pi}\hspace{0.03cm} e^{-2 r^2} \Bigg[\sum_{n=0}^{\infty} (-1)^n L_n^{(0)} (4 r^2) \hspace{0.05cm}\rho_{n,n} + \sum_{j=1}^{\infty} (2 r)^j \sum_{n=0}^{\infty} (-1)^n \frac{L_n^{(j)}(4 r^2)}{\sqrt{(n+1)_j}} \big(e^{-\text{i}j\phi} \rho_{n+j,n}+ e^{\text{i}j\phi} \rho_{n,n+j}\big)\Bigg]\;,
\label{wignermapradial}
\end{equation}
where $(x)_j$ is the Pochhammer symbol, $(x)_j = x(x+1)\dots(x+j-1)$. We see the Wigner function has no angular variation for a purely diagonal density matrix, e.g., in the steady state of an undriven vdP oscillator. Further, elements in the $j$-th off diagonal contribute $j$ units of angular momentum. Equation~\eqref{wignermapradial} can be inverted using orthogonality properties of Laguerre polynomials and exponential functions to give $\rho_{n,n^{\prime}}$ in terms of $W(r,\phi)$,
\begin{equation}
\rho_{n,n^{\prime}} = (-1)^{n^{\prime}} \sqrt{\frac{n^{\prime}!}{n!}} \int_0^{2\pi}\hspace{-0.12cm} d\phi\hspace{0.05cm} e^{\text{i}(n-n^{\prime})\phi}\hspace{-0.03cm} \int_0^{\infty}\hspace{-0.12cm} dr\hspace{0.05cm} e^{-2r^2} (2 r)^{n-n^{\prime}+1}\hspace{0.03cm} L_{n^{\prime}}^{(n-n^{\prime})}(4 r^2)\hspace{0.05cm} W(r,\phi)\;.
\label{wignerinversemap}
\end{equation}

\section{\label{snrsection}Signal-to-noise ratio}
\begin{figure}[h]
\includegraphics[height=0.28\columnwidth]{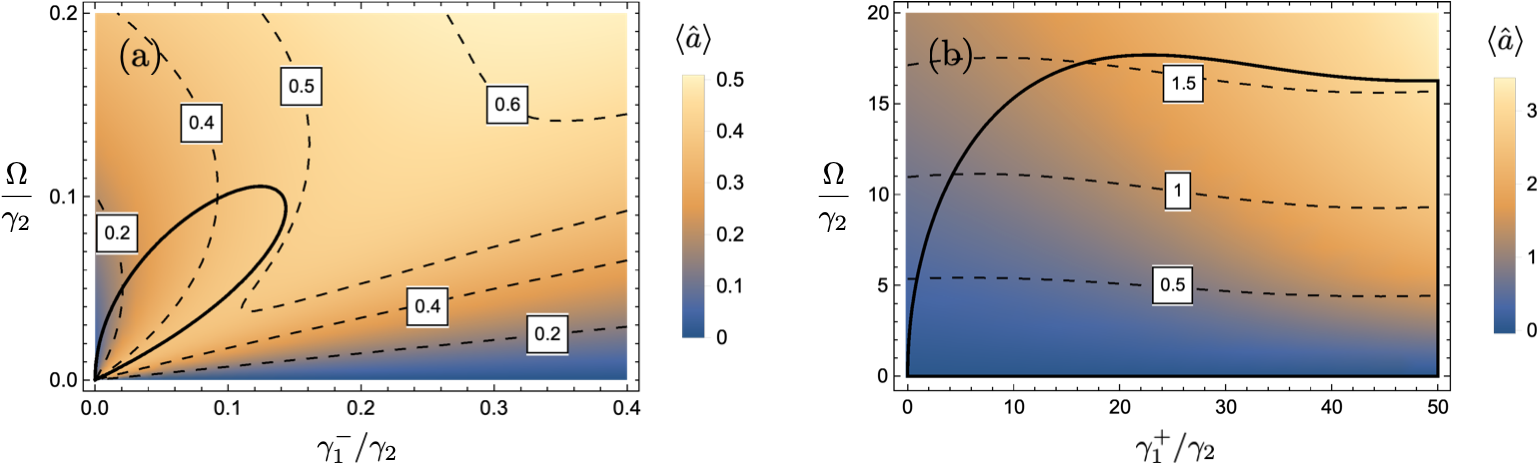}
\caption{\label{snrfig}Response $\langle\hat{a}\rangle$ as a function of drive $\Omega$ and dissipation rates $\gamma_1^{\pm}$. Dashed lines show contours of constant signal-to-noise ratio. (a) $\gamma_1^ + = 0$. Solid curve encloses the region over which $\text{d}\langle\hat{a}\rangle/\text{d}\Omega<0$. (b) $\gamma_1^-/\gamma_2 = 30$. Below the solid curve, one finds a sensitivity enhancement over a passive system, i.e., $\text{d}\langle\hat{a}\rangle/\text{d}\Omega>2/\gamma_1^-$.}
\end{figure}

We have focused on the response $\langle\hat{a}\rangle$ of a vdP oscillator to an external drive. An experimental measurement of this response will be subject to quantum noise. Here we estimate the signal-to-noise ratio. As we showed in Sec.~\ref{wignersection}, the ``signal'' $\langle\hat{a}\rangle$ as well as other physical quantities can be extracted from the Wigner function $W(\alpha)$ which represents a quasiprobability distribution in phase space. From Eq.~\eqref{wignercom}, $\langle\hat{a}\rangle$ is given by the center of mass of $W(\alpha)$. The noise $\sigma$ can be estimated from the spread about this center of mass,
\begin{equation}
\sigma^2 = \int\hspace{-0.05cm} d^2\alpha\hspace{0.05cm} |\alpha - \langle\hat{a}\rangle|^2 \hspace{0.03cm}W(\alpha) =  - |\langle\hat{a}\rangle|^2 + \int\hspace{-0.05cm} d^2\alpha\hspace{0.05cm} |\alpha|^2 \hspace{0.03cm}W(\alpha)\;,
\end{equation}
where we have used Eqs.~\eqref{wignernorm} and \eqref{wignercom}. The last integral can be evaluated from Eq.~\eqref{ensembleavg}, yielding
\begin{equation}
\sigma = \sqrt{\langle\hat{a}^{\dagger} \hat{a} + \hat{a} \hat{a}^{\dagger}\rangle/2 - |\langle\hat{a}\rangle|^2} =  \sqrt{\langle\hat{n}\rangle + 1/2 - |\langle\hat{a}\rangle|^2}\;,
\end{equation}
where $\hat{n}$ is the number operator, $\hat{n} \equiv \hat{a}^{\dagger} \hat{a}$. The signal-to-noise ratio is given by $\text{SNR} = \langle\hat{a}\rangle/\sigma$. We plot the response and the SNR as a function of drive and dissipation in Fig.~\ref{snrfig}, which shows one can have $\text{SNR}\gtrsim 1$ while being in the quantum regime, $\langle\hat{a}\rangle \sim 1$. In Fig.~\ref{snrfig}(a), we consider the case $\gamma_1^{\pm} \ll \gamma_2$, where one finds a nonmonotonic response with increasing drive. Here $\text{SNR} < 1$ as the oscillator is close to the vacuum state. Nonetheless, such states have been measured in experiments with close to 1\% accuracy \cite{sBanaszek1999, sLvovsky2001, sBertet2002, shofheinz2009synthesizing, sChu2018, sSmithey1993, sLeibfried1996}. In Fig.~\ref{snrfig}(b), we find, for $\gamma_1^{\pm} \gg \gamma_2$, one can have both $\text{SNR} > 1$ as well as a susceptibility boost over a passive oscillator, $\chi>\chi_p$.

These noise estimates also show the advantage of measuring weak signals with a quantum vdP oscillator as opposed to simply using the classical limit of the critical oscillator. Although the susceptibility can be made infinitely large in this limit, the particle-number fluctuations also diverge as $\Delta n \sim (\gamma_1^{\pm}/\gamma_2)^{1/2}$ [see Eq.~\eqref{gaussian_crit}]. Therefore, the SNR, in fact, vanishes for weak signals. Instead, by harnessing a quantum oscillator in the limit-cycle regime, one can enhance the ``minimum detectable signal per unit time,'' which is an important figure-of-merit for sensors \cite{sDegen2017}.

%

\end{document}